\documentclass[12pt]{article}
\usepackage{amsfonts,bm}
\usepackage{graphicx}
\usepackage{color}
\baselineskip
\offinterlineskip
\begin{document}

\begin{center}
{\bf Matrices, Fermi Operators and Applications}
\end{center}

\begin{center}
{\bf  Yorick Hardy$^\ast$, Willi-Hans Steeb$^\dag$ and
Garry Kemp$^\dag$} \\[2ex]

$\ast$
School of Mathematics, University of the Witwatersrand, \\
Johannesburg, Private Bag 3, Wits 2050, South Africa, \\
e-mail: {\tt yorick.hardy@wits.ac.za}\\[2ex]

$\dag$
International School for Scientific Computing, \\
University of Johannesburg, Auckland Park 2006, South Africa, \\
e-mail: {\tt steebwilli@gmail.com},
{\tt gkemp@uj.ac.za}\\[2ex]
\end{center}

\strut\hfill

{\bf Abstract.} We consider the vector space
of $n \times n$ matrices over $\mathbb C$, 
Fermi operators and operators constructed from these 
matrices and the Fermi operators. The properties of these 
operators are studied with respect to the underlying matrices.
The commutators, anticommutators, and the eigenvalue problem
of such operators are also discussed.  
Other matrix functions such as the exponential function
are studied. Density operators and Kraus operators are
also discussed.

\strut\hfill

\section{Introduction}

We consider the vector space of $n \times n$ matrices
over $\mathbb C$. Let $c_1^\dagger$, $\cdots$, $c_n^\dagger$,
$c_1$, $\cdots$, $c_n$ be Fermi creation and annihilation
operators, respectively. The anticommutation relations are
(see, for example, \cite{1})
\begin{equation}
\label{eq:1}
[c_j^\dagger,c_k]_+ = \delta_{jk}I, \quad
[c_j,c_k]_+ = 0, \quad [c_j^\dagger,c_k^\dagger]_+ = 0
\end{equation}
where $I$ is the identity operator and 0 the zero operator.
Hence $c_j^2=0$ and $(c_j^\dagger)^2=0$.
Furthermore we have the commutator
\begin{equation}
\label{eq:2}
[c_j^\dagger c_k,c_{\ell}^\dagger c_m] = 
\delta_{k\ell}c_j^\dagger c_m - \delta_{jm}c_{\ell}^\dagger c_k, \quad
j,k,\ell,m=1,\dots,n.
\end{equation}
Let $A$ be an $n \times n$ matrix over $\mathbb C$. 
Then we can form the operator $\hat A$ defined by
\begin{equation}
\label{eq:quad}
\hat A := \pmatrix { c_1^\dagger & \cdots & c_n^\dagger }
A \pmatrix { c_1 \cr \vdots \cr c_n } =
\sum_{j,k=1}^n a_{jk}c_j^\dagger c_k,
\end{equation}
i.e. a quadratic form in the Fermi operators. See \cite{6}
for a motivation of the study of quadratic forms in Fermi operators.
The connection between the quadratic form in (\ref{eq:quad})
and matrices permits the use of numerous techniques and results
from matrix theory \cite{7}.
For example, consider the Pauli spin matrices
$\sigma_1$, $\sigma_2$, $\sigma_3$. Then we have
$$
\hat \sigma_1 = c_1^\dagger c_2 + c_2^\dagger c_1, \quad
\hat \sigma_2 = -ic_1^\dagger c_2 + ic_2^\dagger c_1, \quad
\hat \sigma_3 = c_1^\dagger c_1 - c_2^\dagger c_2.
$$
and 
$$
(\hat \sigma_1)^2 = (\hat \sigma_2)^2 = (\hat \sigma_3)^2 = 
c_1^\dagger c_1 + c_2^\dagger c_2 - 2c_1^\dagger c_1 c_2^\dagger c_2.
$$
We study the properties of the operator $\hat A$ for given matrix 
$A$. We consider normal, nonnormal, hermitian, unitary, 
density matrices etc. The commutator and anticommutator
of two matrices $A$, $B$ and the corresponding operators
\smash{$\hat A$} and \smash{$\hat B$} are also investigated. 

\section{Properties of Matrices} 

Let $I_n$ be the $n \times n$ identity matrix. Then
$$
\hat N = \pmatrix { c_1^\dagger & \cdots & c_n^\dagger }
I_n \pmatrix { c_1 \cr \vdots \cr c_n } =
\sum_{j=1}^n c_j^\dagger c_j
$$
is the number operator. We note that
$$
[c_j^\dagger c_k,c_j^\dagger c_j+c_k^\dagger c_k] = 0.
$$
Let $A$ be an normal matrix, i.e. $AA^*=A^*A$. Then
$\hat A$ is a normal operator. 
Let $A$ be a nonnormal matrix, i.e. $AA^* \ne A^*A$.
Then $\hat A$ is a nonnormal operator. An example 
for $n=2$ is the nonnormal matrix
$$
A = \pmatrix { 0 & 1 \cr 0 & 0 }.
$$
Then we obtain the operator $\hat A=c_1^\dagger c_2$
and hence $\hat A^\dagger=c_2^\dagger c_1$ with 
$$
\hat A\hat A^\dagger=c_1^\dagger c_1(I-c_2^\dagger c_2),
\quad
\hat A^\dagger \hat A=c_2^\dagger c_2(I-c_1^\dagger c_1)
$$
If $H$ is a hermitian matrix, then $\hat H$ is a self-adjoint 
operator. If $C$ is a skew-hermitian matrix, then $\hat C$ is
a skew-hermitian operator. Let $U$ be a unitary matrix. Then
we cannot conclude that $\hat U$ is a unitary operator. As an
example consider 
$$
U = \pmatrix { 0 & 1 \cr 1 & 0 }.
$$
Then $\hat U=c_1^\dagger c_2+c_2^\dagger c_1$, 
$\hat U=\hat U^\dagger$ and     
$$
\hat U\hat U^\dagger = c_1^\dagger c_1 + c_2^\dagger c_2 
-2c_1^\dagger c_1 c_2^\dagger c_2 = 
\hat N-2c_1^\dagger c_1 c_2^\dagger c_2. 
$$
Next we mention that applying the anticommutation relations we obtain
\begin{eqnarray*}
\pmatrix { c_1 \cr c_2 \cr \vdots \cr c_n }
\pmatrix { c_1^\dagger & c_2^\dagger & \cdots & c_n^\dagger }
&=& \pmatrix { c_1c_1^\dagger & c_1c_2^\dagger & \cdots & c_1c_n^\dagger \cr
c_2c_1^\dagger & c_2c_2^\dagger & \cdots & c_2c_n^\dagger \cr
\vdots & \vdots & \ddots & \vdots \cr
c_nc_1^\dagger & c_nc_2^\dagger & \cdots & c_nc_n^\dagger } \\
&=& \pmatrix { I & 0 & \cdots & 0 \cr 0 & I & \cdots & 0 \cr
\vdots & \vdots & \ddots & \vdots \cr 0 & 0 & \cdots & I } - 
\pmatrix { c_1^\dagger c_1 & c_2^\dagger c_1 & \cdots & c_n^\dagger c_1 \cr 
c_1^\dagger c_2 & c_2^\dagger c_2 & \cdots & c_n^\dagger c_2 \cr 
\vdots & \vdots & \ddots & \vdots \cr 
c_1^\dagger c_n & c_2^\dagger c_n & \cdots & c_n^\dagger c_n }.
\end{eqnarray*}
Hence we cannot expect that $(\hat A)(\hat A)=\widehat{(A^2)}$
in general.
Let $A$ be a $2 \times 2$ matrix. Then we obtain
$$
\hat A\hat A = \pmatrix { c_1^\dagger & c_2^\dagger }
A^2 \pmatrix { c_1 \cr c_2 } + 2 \det(A)c_1^\dagger c_1c_2^\dagger c_2.
$$
Thus if $\det(A)=0$, then $\hat A\hat A=\widehat{(A^2)}$.
For a $3 \times 3$ matrix we obtain
$$
(\hat A)^2 = \widehat{(A^2)} 
-2\sum_{{ i<k} \atop {i=1}}^3 \sum_{{ j<\ell} \atop {j=1}}^3 
c_i^\dagger c_k^\dagger c_j c_{\ell}
\det \pmatrix { a_{ij} & a_{i\ell} \cr a_{kj} & a_{k\ell} }. 
$$
Let $A$, $B$ be $n \times n$ matrices. Then we have
\begin{equation}
\label{eq:prod}
\hat A \hat B =
\widehat{AB} -
\sum_{j,k,\ell,m=1 \atop \ell>j,m>k}^n
g\left(\sigma_2 \pmatrix { a_{jk} & a_{jm} \cr a_{\ell k} & a_{\ell m} }
       \sigma_2 \pmatrix { b_{jk} & b_{jm} \cr b_{\ell k} & b_{\ell m} }^T \right)
c_j^\dagger c_{\ell}^\dagger c_k c_m.
\end{equation}
The coefficients are given in terms of the symmetric indefinite form
\begin{eqnarray*}
g(X,Y) &=& g(Y,X) \\
       &=& \mbox{tr}\left(\sigma_2X\sigma_2Y^T\right) \\ 
       &=& \det(X+Y)-\det(X)-\det(Y)
\end{eqnarray*}
over the $2\times 2$ complex matrices, 
where $\sigma_2$ is the Pauli spin matrix
$$
\sigma_2 = \pmatrix { 0 & -i \cr i & 0 }.
$$
Clearly, $g(A,A)=2\det(A)$. The sum in (\ref{eq:prod})
is taken over $g(X,Y)$ for all corresponding $2\times 2$ submatrices
of $A$ and $B$. With $B=A$ we obtain
$$
(\hat A)^2 = \widehat{A^2} - 2\sum_{j,k,\ell,m=1 \atop \ell>j,m>k}
\det ([A]_{j,l;k,m})
c_j^{\dagger}c_{\ell}^\dagger c_kc_m 
$$
where $[A]_{j,l;k,m}$ is the $2\times 2$ submatrix of $A$ taken
from rows $j$ and $l$ and columns $k$ and $m$. Now $\det ([A]_{j,l;k,m})=0$
for all $j,k,l,m$ only if $A$ is a rank-0 or rank-1 matrix.
Let ${\bf v}$ be a normalized (column) vector in ${\mathbb C}^2$.
Then $\rho={\bf v}{\bf v}^*$ is a density matrix (pure state).
Now 
$$
\hat \rho = \pmatrix { c_1^\dagger & c_2^\dagger }
{\bf v}{\bf v}^* \pmatrix { c_1 \cr c_2 }
= v_1{\overline v}_1c_1^\dagger c_1 +
v_1{\overline v}_2 c_1^\dagger c_2 + 
v_2{\overline v}_1 c_2^\dagger c_1 +
v_2{\overline v}_2 c_2^\dagger c_2.
$$
Obviously $\hat \rho$ is a self-adjoint operator. We have
$\hat \rho=(\hat \rho)^2$. 
Hence $\hat \rho$ is a density operator. Obviously this also
holds for a normalized vector $\bf v$ in ${\mathbb C}^n$ and $\rho={\bf v}{\bf v}^*$,
since $\rho$ is a rank-1 matrix.
\newline

Let $\Pi$ be a projection matrix, i.e. $\Pi=\Pi^*$
and $\Pi=\Pi^2$. Then obviously $\hat \Pi=\hat \Pi^\dagger$
and $\hat \Pi^2 = \hat\Pi$ if and only if $\Pi$ is a
projection onto a one-dimensional subspace (i.e. has rank 1).
We note that $\det(\Pi)=0$ except when $\Pi$ is the identity matrix.
\newline

The symmetry of $g(X,Y)$ yields that
$
[\hat A,\hat B] = \widehat{[A,B]}.
$
Finally, we note that $A\mapsto\hat A$ is trace preserving, i.e.
$$
\mbox{tr}(\hat A) = \sum_{k=1}^n\langle 0|c_k\hat Ac_k^\dagger|0\rangle
  = \mbox{tr}(A)
$$
where $|0\rangle$ is the vacuum state,  $c_k|0\rangle=0$ and
$\langle0|c_k^\dagger=0$.

\section{Exponential Function}

Let $C=(C_{jk})$, $C_1$, $C_2$, 
be $n \times n$ skew-hermitian matrices $(j,k=1,\dots,n)$.
Then $V=\exp(C)$, $V_1=\exp(C_1)$, 
$V_2=\exp(C_2)$ are unitary matrices and
$$
\hat U(V) := \exp(\hat C) = \exp\left(\sum_{j=1}^n\sum_{k=1}^n C_{jk}c_j^\dagger c_k\right)
$$
is a unitary operator with $V=\exp(C)$. 
If $C$ is a rank-1 matrix we have that $\hat{(e^{C})}=e^{\hat C}.$
Owing to the commutator given in equation (\ref{eq:2}) we obtain
\begin{eqnarray*}
\hat U(V_1)\hat U(V_2) &=& \hat U(V_1V_2) \\
\hat U(V^{-1}) &=& \hat U^{-1}(V) = \hat U^\dagger(V) \\
\hat U(I_n) &=& I
\end{eqnarray*}
where $I$ is the identity operator.
\newline

We also note that the Baker-Campbell-Hausdorff formula can be expressed
in terms of repeated commutators \cite{8}
$$
\log(e^Xe^Y) = \sum\frac1{p+q}
               \frac{(-1)^{k-1}}{k}\frac{1}{p_1!q_1!\cdots p_k!q_k!}
 [X^{p_1}Y^{q_1}\cdots X^{p_k}Y^{q_k}]
$$
where the sum is taken over all $k\in\mathbb{N}$ and $p,q,p_1,q_1,\ldots,p_k,q_k\in\mathbb{N}_0$
such that
$$
\sum_{i=1}^kp_i = p, \quad \sum_{i=1}^kq_i = q, \quad p_i+q_i>0,\;\; i\in\{1,2,\ldots,k\}
$$
and the repeated commutators are given by
$$
[X^{p_1}Y^{q_1}\cdots X^{p_k}Y^{q_k}]
:=        [\,\overbrace{X\cdots,[X}^{p_1}
          ,[\,\overbrace{Y\cdots,[Y}^{q_1}
           ,[X\cdots,[Y,[\,\overbrace{X\cdots,[X}^{p_k}
            ,[\,\overbrace{Y\cdots,[Y,Y}^{q_k}]\cdots].
$$
Since $[\hat A,\hat B] = \widehat{[A,B]}$, we have
$$
\widehat{\log(e^Xe^Y)} = \log(e^{\hat X}e^{\hat Y}).
$$

\section{Commutators, Lie Algebras and Anticommutators}

Let $A$, $B$ be $n \times n$ matrices and
$\hat A$, $\hat B$ be the corresponding operators.
Then a straightforward calculation shows that
$$
[\hat A,\hat B] = 
\pmatrix { c_1^\dagger & \cdots & c_n^\dagger }
[A,B]
\pmatrix { c_1 \cr \vdots \cr c_n } = 
\widehat{[A,B]}
$$
As an example consider the $2 \times 2$ matrices
$$
A = \pmatrix { 0 & 1 \cr 0 & 0 }, \qquad
B = \pmatrix { 0 & 0 \cr 0 & 1 }.
$$
Then $[A,B]=A$ and $A$, $B$ form a basis of a 
two-dimensional non-abelian Lie algebra. Now
$$
\hat A = c_1^\dagger c_2, \qquad
\hat B = c_2^\dagger c_2
$$
and 
$$
[\hat A,\hat B] = c_1^\dagger c_2 = \hat A.
$$
Hence the operators $\hat A$, $\hat B$ form a basis
of a two-dimensional non-abelian Lie algebra.
For two arbitrary $2 \times 2$ matrices $A$, $B$ we have
$$
[A,B] = 
\pmatrix { a_{12}b_{21}-a_{21}b_{12} & 
\mbox{tr}(\sigma_3A)b_{12} - \mbox{tr}(\sigma_3B)a_{12} \cr
-\mbox{tr}(\sigma_3A)b_{21} + \mbox{tr}(\sigma_3B)a_{21} & 
a_{21}b_{12} - b_{21}a_{12} }.
$$
Now 
\begin{eqnarray*}
\hat A &=& 
a_{11}c_1^\dagger c_1 + a_{12}c_1^\dagger c_2 + 
a_{21}c_2^\dagger c_1 + a_{22}c_2^\dagger c_2 \\
\hat B &=& 
b_{11}c_1^\dagger c_1 + b_{12}c_1^\dagger c_2 + 
b_{21}c_2^\dagger c_1 + b_{22}c_2^\dagger c_2.
\end{eqnarray*}
Then
\begin{eqnarray*}
[\hat A,\hat B] &=& 
(c_1^\dagger c_1-c_2^\dagger c_2)(a_{12}b_{21}-a_{21}b_{12}) \\
&& + (a_{11}b_{12}-a_{12}b_{11}+a_{12}b_{22}-a_{22}b_{12})c_1^\dagger c_2 
+ (a_{21}b_{11}-a_{21}b_{22}+a_{22}b_{21}-a_{11}b_{21})c_2^\dagger c_1.
\end{eqnarray*}
Investigating the anticommutator between $\hat{A}$ and $\hat{B}$, we find
$$
	[\hat{A},\hat{B}]_{+} = \widehat{[A,B]_{+}} - 2\sum_{j,k,\ell,m=1 \atop \ell>j,m>k}^n g\left( [A]_{j,l;k,m} ,  [B]_{j,l;k,m}  \right) c^{\dagger}_{j} c^{\dagger}_{l} c_{k} c_{m}
$$
where $[A]_{j,l;k,m} := \pmatrix { a_{jk} & a_{jm} \cr a_{\ell k} & a_{\ell m} }.$
For example, with $n = 2$,
$$
	[\hat{A},\hat{B}]_{+} = \widehat{[A,B]_{+}} + 2 \left(a_{11}b_{22} + a_{22}b_{11} - a_{12}b_{21} - a_{21}b_{12}\right) c^{\dagger}_{1} c_{1} c^{\dagger}_{2} c_{2},
$$
with
$$
	[A,B]_{+} = 
\pmatrix { 2a_{11}b_{11}+a_{12}b_{21} + a_{21}b_{12} & 
\mbox{tr}\left( A \right)b_{12} + \mbox{tr}\left( B \right)a_{12} \cr
\mbox{tr}\left( A \right)b_{21} + \mbox{tr}\left( B \right)a_{21} & 
 2a_{22}b_{22}+a_{21}b_{12} + a_{12}b_{21} }.
$$

Consider the Pauli spin matrices. Then we have
$$
[\sigma_1,\sigma_2]_+ = 0_2, \quad
[\sigma_2,\sigma_3]_+ = 0_2, \quad
[\sigma_3,\sigma_1]_+ = 0_2
$$
and
$$
[\hat \sigma_1,\hat \sigma_2]_+ = 0, \quad
[\hat \sigma_2,\hat \sigma_3]_+ = 0, \quad
[\hat \sigma_3,\hat \sigma_1]_+ = 0.
$$
It follows that
$$
	[A,B]_{+} = 
\mbox{tr}(A)B+\mbox{tr}(B)A
 -\frac12\left(\mbox{tr}(A)\mbox{tr}(B)
              -\sum_{j=1}^3\mbox{tr}(\sigma_jA)\mbox{tr}(\sigma_jB)
         \right)I_2.
$$

\section{Eigenvalue Problem} 

Consider first the $2 \times 2$ case with
$$
\hat A = a_{11}c_1^\dagger c_1 + a_{12}c_1^\dagger c_2 +
a_{21}c_2^\dagger c_1 + a_{22}c_2^\dagger c_2.
$$
With the basis element $|0\rangle$ and $c_j|0\rangle=0|0\rangle$
we obtain $\langle 0|\hat A|0\rangle=0$. 
With the basis element $c_2^\dagger c_1^\dagger|0\rangle$
we obtain 
$$
\hat A c_2^\dagger c_1^\dagger|0\rangle =
a_{11}c_2^\dagger c_1^\dagger|0\rangle +
a_{22}c_2^\dagger c_1^\dagger|0\rangle =
(a_{11}+a_{22})c_2^\dagger c_1^\dagger|0\rangle =
\mbox{tr}(A)c_2^\dagger c_1^\dagger|0\rangle.
$$
With the basis $c_1^\dagger|0\rangle$, $c_2^\dagger|0\rangle$
we obtain 
$$
\hat Ac_1^\dagger|0\rangle = 
a_{11}c_1^\dagger|0\rangle + a_{21}c_2^\dagger|0\rangle
$$
$$
\hat Ac_1^\dagger|0\rangle = 
a_{11}c_1^\dagger|0\rangle + a_{21}c_2^\dagger|0\rangle
$$
Together with the dual basis $\langle 0|c_1$, $\langle 0|c_2$
we obtain that the matrix representation of $\hat A$
is $A$.
\newline

For the $3 \times 3$ case we have
$$
\hat A = \sum_{j,k=1}^3 a_{jk}c_j^\dagger c_k.
$$
With the basis $|0\rangle$ we obtain $\langle 0|\hat A|0\rangle=0$. 
For the basis $c_1^\dagger|0\rangle$, $c_2^\dagger|0\rangle$, 
$c_3^\dagger|0\rangle$ and the respectively dual basis
$\langle 0|c_1$, $\langle 0|c_2$, $\langle 0|c_3$
we obviously obtain $A$ as the matrix representation
of $\hat A$. With the basis 
$$
c_1^\dagger c_2^\dagger |0\rangle, \quad
c_1^\dagger c_3^\dagger |0\rangle, \quad 
c_2^\dagger c_3^\dagger |0\rangle
$$
and the corresponding dual one
$$
\langle 0|c_2c_1, \quad 
\langle 0|c_3c_1, \quad 
\langle 0|c_3c_2
$$
we obtain the matrix representation of $\hat A$
$$
\pmatrix { a_{11}+a_{22} & a_{23} & -a_{13} \cr
a_{32} & a_{11}+a_{33} & a_{12} \cr
-a_{31} & a_{21} & a_{22}+a_{33} }.
$$
Note that the trace of this matrix is twice the trace of $A$.
In general we have
$$
\hat A = \sum_{j,k=1}^n a_{j,k}c_j^\dagger c_k
$$
and with the basis 
$$
\{ \, c_j^\dagger|0\rangle \,:\, j=1,\dots,n \, \}
$$
the matrix representation of $\hat A$ is given by $A$,
and the matrix representation of $\hat A\hat B$ is $AB$.
The trace of $A$ is an eigenvalue of $\hat A$:
$$
\hat A c_n^\dagger c_{n-1}^\dagger\cdots c_1^\dagger|0\rangle
  = \mbox{tr}(A) c_n^\dagger c_{n-1}^\dagger\cdots c_1^\dagger|0\rangle.
$$

\section{Kraus Operators}

Consider the Kraus operators $K_1$ and $K_2$
$$
K_1 = \pmatrix { 0 & 1 \cr 0 & 0 } \,\,\, \Rightarrow \,\, K_1^* = \pmatrix { 0 & 0 \cr 1 & 0 }, \quad
K_2 = \pmatrix { 0 & 0 \cr 1 & 0 } \,\,\, \Rightarrow \,\, K_2^* = \pmatrix { 0 & 1 \cr 0 & 0 }
$$
and an arbitrary $2 \times 2$ matrix $A=(a_{jk})$. Then 
$$
K_1AK_1^* + K_2AK_2^* = \pmatrix { a_{22} & 0 \cr 0 & a_{11} }.
$$
So the trace of $A$ is preserved under this transformation.
Let $c_1^\dagger$, $c_2^\dagger$, $c_1$, $c_2$ be Fermi creation and annihilation
operators, respectively. Then 
$$
\hat K_1 = \pmatrix { c_1^\dagger & c_2^\dagger }\pmatrix { 0 & 1 \cr 0 & 0 }
\pmatrix { c_1 \cr c_2 } =
c_1^\dagger c_2, \quad \hat K_1^\dagger = c_2^\dagger c_1
$$
$$
\hat K_2 = \pmatrix { c_1^\dagger & c_2^\dagger }\pmatrix { 0 & 0 \cr 1 & 0 }
\pmatrix { c_1 \cr c_2 } =
c_2^\dagger c_1, \quad \hat K_2^\dagger = c_1^\dagger c_2
$$
and 
$$
\hat A = \pmatrix { c_1^\dagger & c_2^\dagger } A \pmatrix { c_1 \cr c_2 } =
a_{11}c_1^\dagger c_1 + a_{12}c_1^\dagger c_2 + a_{21}c_2^\dagger c_1 
+ a_{22}c_2^\dagger c_2.
$$
It follows that 
$$
\hat K_1\hat A\hat K_1^\dagger + \hat K_2 \hat A\hat K_2^\dagger =
a_{22}c_1^\dagger c_1 + a_{11} c_2^\dagger c_2
 + (a_{11} + a_{22}) c_1^\dagger c_2^\dagger c_1c_2.
$$
Thus the embedding preserves this map when $\mbox{tr}(A)=0.$
Let $K_1$, \ldots, $K_r$ be matrix Kraus operators.
As we noted in the previous section, in the basis
$$
\{ \, c_j^\dagger|0\rangle \,:\, j=1,\dots,n \, \}
$$
the matrix representation of $\hat A$ is given by $A$,
and the matrix representation of
$
\hat K_1\hat A\hat K_1^* + \cdots + \hat K_r\hat A\hat K_r^*
$
is precisely
$
K_1AK_1^* + \cdots + K_rAK_r^*.
$

\section{Extensions}

In order to model all quadratic forms in the Fermi operators
we need to add operators of the form
$$
\hat B = \pmatrix { c_1^\dagger & c_2^\dagger & \cdots & c_n^\dagger }
B 
\pmatrix { c_1^\dagger \cr c_2^\dagger \cr \vdots \cr c_n^\dagger }
$$
and 
$$
\hat D = 
\pmatrix { c_1 & c_2 & \cdots & c_n } D 
\pmatrix { c_1 \cr c_2 \cr \vdots \cr c_n }
$$
where we have to take into account that $c_j^2=0$, 
$(c_j^\dagger)^2=0$. For $n=2$ we have
$$
\hat B = (b_{12}-b_{21})c_1^\dagger c_2^\dagger
$$
$$
\hat D = (d_{12}-d_{21})c_1c_2.
$$
Now the commutator of $\hat B$ and $\hat D$ is given by
$$
[\hat B,\hat D] = (b_{12}-b_{21})(d_{12}-d_{21})(I-c_1^\dagger c_1-c_2^\dagger c_2). 
$$
Finally, we may consider Fermi-Bose coupled quadratic forms
$$
\hat M = 
\left(
\pmatrix { c_1^\dagger & c_2^\dagger & \cdots & c_n^\dagger }
\otimes
\pmatrix { b_1^\dagger & b_2^\dagger & \cdots & b_m^\dagger }
\right)
M 
\left(
\pmatrix { c_1 \cr c_2 \cr \vdots \cr c_n }
\otimes
\pmatrix { b_1 \cr b_2 \cr \vdots \cr b_m }
\right).
$$
A decomposition
$$
M = \sum_{j=1}^r M_{c,j}\otimes M_{b,j}
$$
of the $mn\times mn$ matrix $M$ over the $n\times n$ matrices $M_{c,r}$
and $m\times m$ matrices $M_{b,r}$ yields a sum of quadratic forms in
the Fermi operators coupled with quadratic forms in the Bose operators.
However, we no longer have the straightforward relationship between the matrix
commutator and the commutator of Fermi/Bose operators.

\strut\hfill

\end{document}